\documentclass[prl,twocolumn,amsmath,amssymb]{revtex4}

\usepackage{graphicx}
\usepackage{dcolumn}
\usepackage{bm}
\usepackage[ansinew]{inputenc}
\usepackage[T1]{fontenc}
\usepackage{color}
\usepackage{ae,aecompl}
\usepackage{textcomp}

\begin{document}

\title{Lossless State Detection of Single Neutral Atoms}

\author{J.~Bochmann}
\author{M.~M{\"{u}}cke}
\author{C.~Guhl}
\author{S.~Ritter}
\email{stephan.ritter@mpq.mpg.de}
\author{G.~Rempe}
\author{D.~L.~Moehring}
\altaffiliation[Present address: ]
{Sandia National Laboratories, Albuquerque, NM 87185, USA}
\email{dlmoehr@sandia.gov}

\affiliation{Max-Planck-Institut f{\"{u}}r Quantenoptik, Hans-Kopfermann-Strasse 1, 85748 Garching, Germany}


\begin{abstract}
We introduce lossless state detection of trapped neutral atoms based on cavity-enhanced fluorescence. In an experiment with a single $^{87}$Rb atom, a hyperfine-state-detection fidelity of 99.4\,\% is achieved in 85\,\textmu s. The quantum bit is interrogated many hundreds of times without loss of the atom while a result is obtained in every readout attempt. The fidelity proves robust against atomic frequency shifts induced by the trapping potential. Our scheme does not require strong coupling between the atom and cavity and can be generalized to other systems with an optically accessible quantum bit.
\end{abstract}


\maketitle

Current efforts in experimental physics aim at gaining control over fundamental quantum systems. Single neutral atoms are a prime example, reflected in groundbreaking work on feedback control, quantum transport, gate operations, and entanglement \cite{kimble:2008, kubanek:2009, meschede:2009, grangier:2010, saffman:2010}. While optical fields directed at the atoms provide excellent control of atomic states, the retrieval of information about the internal state of a single atom is difficult. In this Letter, we introduce a controlled readout channel by coupling a single atom to an optical cavity. The cavity enhances the matter-light interaction and allows efficient detection of the internal atomic state. The feasibility of high-fidelity state detection without loss of the atom establishes single neutral atoms as truly stationary carriers of quantum information.

In a single atom, quantum information is typically encoded in or mapped onto electronic hyperfine states which can be spectroscopically resolved in fluorescence measurements. State readout based on fluorescence light detection distinguishes two atomic states by detecting either a high rate of scattered photons (identified as ``bright'' state) or no scattered photons (``dark'' state) when the atom is state-selectively excited with a probe laser. This method is the most powerful technique today \cite{blatt:2004, acton:2006, hume:2007, olmschenk:2007} and has been employed in all quantum computing protocols with single ions in Paul traps. Recent experiments report readout fidelities as high as 99.99\,\% with a single trapped calcium ion \cite{myerson:2008}. But despite its success, fluorescence state detection alone has never been realized with a trapped neutral atom. This is due to the difficulty to detect a sufficient number of scattered photons from an atom in the bright state before it is ejected from the trap. This state-dependent loss of the atom has effectively been used for state detection. In such pushout schemes \cite{kuhr:2005, volz:2006, yavuz:2006, jones:2007, lengwenus:2007}, the loss of the atom signals one of the internal states.

Here, we demonstrate lossless hyperfine-state readout of a single trapped $^{87}$Rb atom based on cavity-enhanced fluorescence. The atomic state can be interrogated many hundred times without loss of the atom from an optical dipole trap. We achieve a hyperfine state-detection fidelity of 99.4\,\% in 85\,\textmu s while an answer is obtained in every readout attempt. In contrast to previous attempts on cavity-assisted readout \cite{reichel:2010, khudaverdyan:2009, boozer:2006}, we do not require the strong-coupling regime of cavity QED or ground state cooled atoms, which facilitates implementation in a wide range of physical systems.
\begin{figure}[t]
\includegraphics[width=1.0\columnwidth,keepaspectratio]{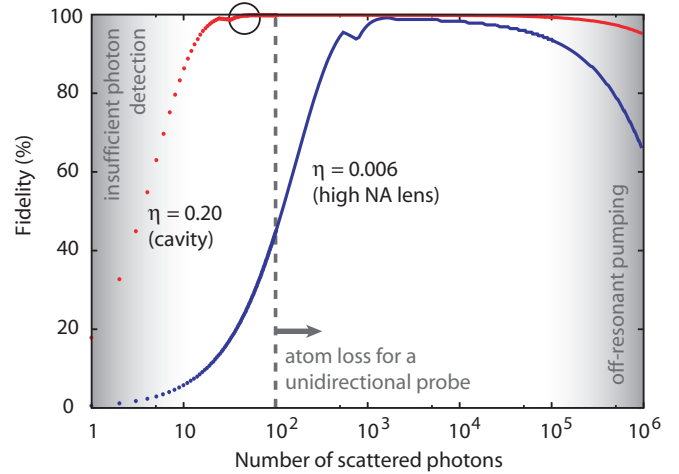}
\caption{\label{fig:fidelities}
Calculated maximum fidelities for fluorescence state detection of a single $^{87}$Rb atom with a cavity ($\eta$=20\,\%) and with a high numerical aperture (NA) objective ($\eta$=0.6\,\%), where $\eta$ is the detection probability of a scattered photon. Fidelities are limited by an insufficient number of detected photons, optical pumping, and detector dark counts. Unidirectional probe light expels the atom from an optical dipole trap after scattering $\approx100$ photons (dashed line) for a typical trap depth of 2\,mK. The novel regime of cavity-enhanced fluorescence readout introduced in this Letter is indicated by a black circle. A state detection fidelity of 99.98\,\% is feasible with less than 100 scattered photons.}
\end{figure}

To appreciate the crucial role of the photon detection efficiency in fluorescence state detection, we briefly analyze two detection scenarios. We contrast the achievable state-detection fidelity for fluorescence photon collection with a high numerical aperture objective and with an optical cavity (Fig.~\ref{fig:fidelities}). In our example, we choose the $^{87}$Rb  \mbox{5S$_{1/2}~F=2$} state as a bright and the \mbox{5S$_{1/2}~F=1$} state as a dark state. Ideally, a $\sigma^+$-polarized probe laser could drive the \mbox{$\left|F=2,m_F=2\right\rangle\leftrightarrow \left|F'=3,m_{F'}=3\right\rangle$} cycling transition such that off-resonant pumping into the dark hyperfine state is suppressed. In practice, such a unidirectional laser beam quickly ejects the atom from an optical dipole trap before scattering a number of photons sufficient to identify the atomic hyperfine state. For this reason, counter-propagating laser beams are necessary to balance radiation pressure. Applying a lin$\bot$lin polarization configuration avoids standing light wave effects and can even cool the atom during probing \cite{nussmann:2005b}. However, off-resonant excitation of the nearby \mbox{5P$_{3/2}~F'=2$} state opens a decay channel to the dark $F=1$ state. Hence, high-fidelity atomic state readout requires a sufficient number of fluorescence photons to be detected before pumping into the dark state and before atom loss occurs---making the photon detection efficiency a decisive parameter.

Assuming a total photon detection efficiency $\eta$=0.6\,\% with a high numerical aperture objective (best reported value in a single-atom setup, \cite{darquie:2005}), the achievable fidelity is limited to 99.0\,\% and requires scattering of thousands of photons (Fig.~\ref{fig:fidelities}). However, $\eta$ can be dramatically increased with an optical cavity by means of the Purcell effect \cite{orozco:2009, kozuma:2009}. It not only enhances the total fluorescence scattering rate but also channels the photons into a well-defined cavity output mode. This occurs with a rate $2 g^2/\kappa$ where $g$ denotes the coherent atom-cavity coupling constant and $\kappa$ the cavity-field decay rate. For the setup considered in this work, this causes about 60\,\% of all scattered photons to be emitted into the cavity mode resulting in a total detection probability of $\eta$=20\,\% per scattered photon. Moreover, the Purcell-enhancement of the fluorescence transition leads to a relative suppression of off-resonant decay paths and therefore reduces the effect of unwanted bright to dark state pumping. All in all, a remarkably high atomic state readout fidelity of 99.98\,\% can be achieved with less than 100 scattered photons (Fig.~\ref{fig:fidelities}).

The calculation of achievable fidelities (Fig.~\ref{fig:fidelities}) takes into account the fluorescence photon detection efficiency $\eta$, probe laser induced optical pumping \cite{acton:2006}, the full level scheme of $^{87}$Rb and detector dark counts (25\,s$^{-1}$). The probe laser is assumed to be retroreflected in a lin$\bot$lin polarization configuration with saturation parameter $s$=0.1. We employ a conservative definition of the fidelity as the minimum probability with which the correct atomic hyperfine state is inferred from any photon number detected in a single readout attempt. Nonmonotonic behavior of the achievable fidelity occurs at shifts of the discrimination level between bright and dark state signal due to the trade-off between a sufficient bright state signal and a tolerable number of dark counts \cite{myerson:2008}.

Quantitatively, the rate of scattered photons $R_{\text{scat}}$ at the cavity output scales with the excitation probability $P_{\text{e}}$ of the intracavity atom as
\begin{equation}
	R_{\text{scat}}= 2\kappa \frac{g^{2}}{\Delta_c^{2}+\kappa^{2}} P_{\text{e}},
\end{equation}
where the excitation probability of the atom in free space $P_{\text{e,free}}$ is affected by the presence of the cavity as \mbox{$P_{\text{e}}= P_{\text{e,free}}/(\left|1-\nu\right|^{2})$}. The complex cooperativity $\nu=g^{2}/\left[\left(\Delta_a-i\gamma\right)\left(\Delta_c-i\kappa\right)\right]$ includes the detuning of atom $\Delta_{a}$ and cavity $\Delta_{c}$ with respect to the probe laser \cite{murr:2003}, where $\gamma$ is the atomic polarization decay rate. The scattering rate enhancement only weakly depends on cavity length $l$ in the near-planar limit ($2 g^2/\kappa \propto l^{-1/2}$). This greatly relaxes constraints on cavity parameters and facilitates implementation.
\begin{figure}[b]
\includegraphics[width=1.0\columnwidth,keepaspectratio]{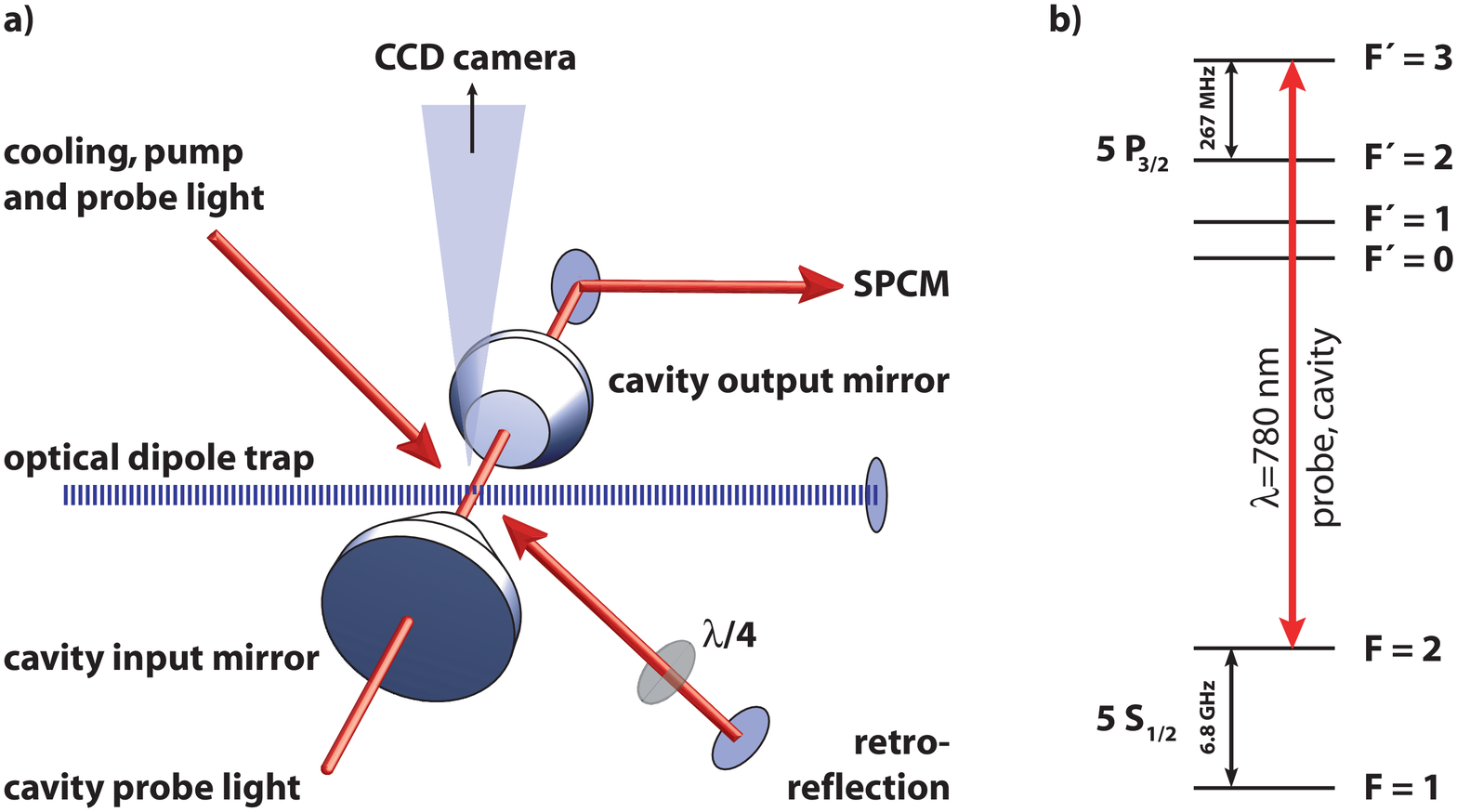}
\caption{\label{fig:setup}
(a) Experimental apparatus. A single $^{87}$Rb atom is trapped in an optical cavity at the focus of a standing-wave dipole trap. A CCD-camera system monitors the position of the atom (inset: CCD-camera image of a single intra-cavity atom, image size 15\,\textmu m $\times$ 25\,\textmu m). For optical cooling and state preparation of the atom, laser beams near resonant with the \mbox{5S$_{1/2}\leftrightarrow$~5P$_{3/2}$} transitions are applied orthogonal to the cavity axis and retroreflected with a lin$\bot$lin polarization ($\lambda /4$: quarter wave-plate). For atomic state detection, a probe laser resonant with the \mbox{$F=2\leftrightarrow$$F'=3$} transition is applied either orthogonal to the cavity axis for fluorescence state detection or along the cavity axis for differential transmission measurements. Photons emitted into the cavity output mode are detected by a single photon counting module (SPCM).
(b) Energy level diagram of the $^{87}$Rb $D_2$-transition, not to scale.
}
\end{figure}

In our experiment (Fig.~\ref{fig:setup}), a single $^{87}$Rb atom is trapped for up to 30 sec at the focus of a standing-wave laser beam (waist radius 16\,\textmu m, power $2.5$\,W, wavelength 1064\,nm, potential depth $2$\,mK, linear polarization) in the center of an optical cavity \cite{bochmann:2008}. The cavity mirrors are separated by 495\,\textmu m with a TEM$_{00}$-mode waist radius of 30\,\textmu m and a finesse of 56000. The cavity is optically asymmetric (mirror transmissions 2\,ppm and 101\,ppm, losses 10\,ppm) such that 90\,\% of the photons inside the cavity exit the resonator through the higher transmission mirror. The cavity output mode is coupled to a single mode optical fiber which is connected to a single photon counting module (quantum efficiency 50\,\%). The total detection efficiency for a photon which has been emitted through the cavity output mirror is 40\,\%. The average atom-cavity coupling for the \mbox{$\left|F=2\right\rangle\leftrightarrow$~$\left|F'=3\right\rangle$} transition is $g_{\text{av}}/2\pi=3$\,MHz, including spatial averaging of $g$ along the cavity axis and over all Clebsch-Gordan coefficients. The cavity and atomic decay rates $\kappa$ and $\gamma$ are $(\kappa,\gamma)/2\pi=(2.8,3.0)$\,MHz, respectively. With a CCD-camera system (numerical aperture 0.4, spatial resolution $1.3$\,\textmu m) we determine the position of single atoms trapped in the cavity by collecting light scattered during optical cooling of the atoms. For the data presented here, we trap exactly one atom in the center ($\pm 10$\,\textmu m) of the cavity mode.

To characterize cavity-enhanced fluorescence state detection, we repeatedly apply a protocol of optical cooling, atomic state preparation and atomic state readout at a rate of 400\,Hz. The atom is first cooled (2\,ms), alternately prepared in the \mbox{$F=1$} or \mbox{$F=2$} hyperfine ground state by optical pumping (100\,\textmu s) and finally probed during a state-detection interval (85\,\textmu s). We set the cavity and probe laser frequencies equal ($\Delta_c=0$) and red-detuned from the \mbox{$F=2\leftrightarrow$$F'=3$} atomic resonance by $\Delta_{a}/2\pi=30$\,MHz, where $\Delta_{a}$ refers to the detuning between probe laser and Stark-shifted atomic transition. The detuning is chosen to avoid probe-light induced heating of the atom. The probe laser is applied orthogonal to the cavity axis and retroreflected in a lin$\bot$lin polarization configuration with a power of $40$\,nW and a beam waist radius of $\approx50$\,\textmu m.

\begin{figure}[htb]
\includegraphics[width=1.0\columnwidth,keepaspectratio]{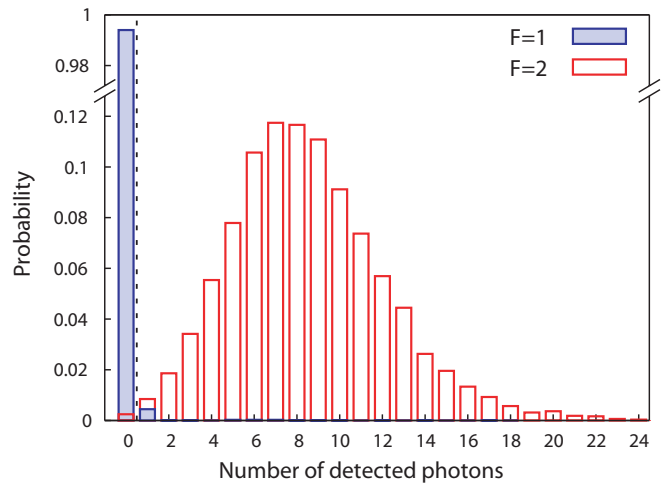}
\caption{\label{fig:fluorescence}
Cavity-enhanced fluorescence state detection. Shown are measured probability distributions for the number of detected photons N per probe interval. The atom is either prepared in the \mbox{$F=2$} (open red histogram) or in the \mbox{$F=1$}  hyperfine ground state (filled blue histogram). The atom is illuminated with a 85\,\textmu s pulse of probe light orthogonal to the cavity axis and resonant with the \mbox{F=2$\leftrightarrow F'=3$} transition. Identifying probe intervals with $N=0$ as the \mbox{$F=1$} state and intervals with $N\geq1$ as the \mbox{$F=2$} state (dashed discrimination line) yields a hyperfine state-detection fidelity of 99.4$\pm$0.1\,\% (uncertainty is statistical), mainly limited by detector dark counts and state preparation errors.
The bright state photon number distribution is nearly Poissonian, but broadened because atomic position and Stark-shift uncertainties lead to shot-to-shot variations of the mean number of scattered photons (Mandel Q parameter Q=0.5).}
\end{figure}

Analyzing the number of detected fluorescence photons $N$ during each probe interval, we find a clear distinction between the dark \mbox{$F=1$} and the bright \mbox{$F=2$} hyperfine state (Fig.~\ref{fig:fluorescence}). Identifying probe intervals with $N=0$ as the \mbox{$F=1$} state and intervals with $N\geq1$ as the \mbox{$F=2$} state results in a hyperfine state detection fidelity of 99.4$\pm$0.1\,\% (uncertainty is statistical). The measured fidelity is limited by state preparation errors (failure of optical pumping, contribution to infidelity $\approx$0.1\,\%) and false counts (electronic dark counts 25\,s$^{-1}$, stray light counts 25\,s$^{-1}$) of the photodetector (contribution to infidelity $\approx$0.4\,\%). In the presented data, the hyperfine state of a single atom was typically interrogated 800 times without loss of the atom.

\begin{figure}[htb]
\includegraphics[width=1.0\columnwidth,keepaspectratio]{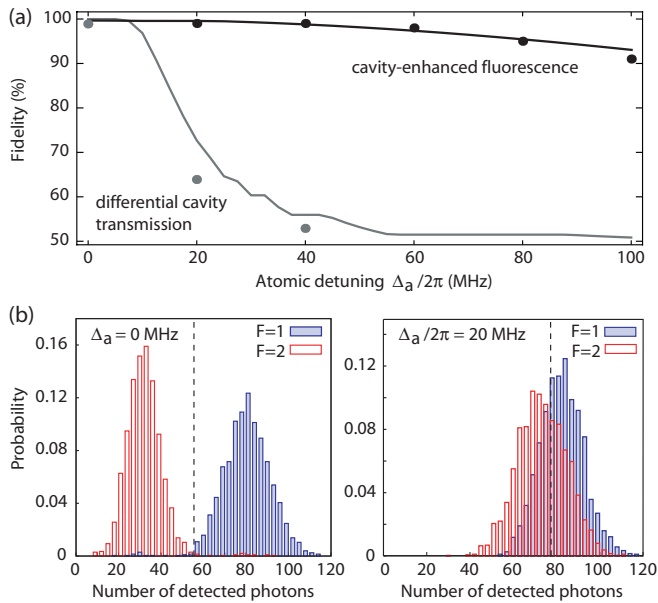}
\caption{\label{fig:detuning}
Fidelity as a function of atomic detuning for state detection by fluorescence and differential transmission.
(a) The state detection fidelity is robust against atomic detuning for the fluorescence technique (black solid line: calculated fidelity for on average N=8 detected photons for the \mbox{$F=2$} state, black dots: measured fidelity). For the differential transmission technique, the fidelity decreases rapidly with detuning (solid grey line: calculated fidelity for on average N=80 detected photons for the \mbox{$F=1$} state, grey dots: measured fidelity). (b) Differential transmission histograms measured at low ($\Delta_a=0$\,MHz) and higher ($\Delta_a/2\pi=20$\,MHz) atomic detuning for 300\,\textmu s transmission probe intervals. The atom is prepared in the \mbox{$F=2$} (open red histogram) or \mbox{$F=1$} state (filled blue histogram). We choose discrimination levels (vertical dashed lines) which optimize the fidelity and measure a state-detection fidelity of 99.0$\pm$0.5\,\% at $\Delta_a=0$\,MHz and a fidelity of 64$\pm$2\,\% at $\Delta_a/2\pi=20$\,MHz (uncertainty is statistical).
}
\end{figure}

Next, we investigate the dependence of state readout fidelity on atomic detuning (Fig.~\ref{fig:detuning}). This is important because neutral atoms are usually trapped in optical dipole potentials and may experience significant ac-Stark shift variations. In our experiment, we mimic this effect by keeping the probe laser and cavity resonant ($\Delta_c=0$) and by detuning them from the atomic resonance (probe-atom detuning $\Delta_a/2\pi=0 ... 100$\,MHz). The probe laser power is increased with $\Delta_a$ such that the mean photon number detected from the bright state is kept constant (N$\approx$8 on average). Fidelities on the order of 99\,\% are maintained up to 40\,MHz detuning, decreasing to 91\,\% at $\Delta_a/2\pi=100$\,MHz due to off-resonant excitation of the \mbox{$F'=2$} state as $\Delta_a$ approaches the excited state hyperfine splitting ($\omega_{\text{HFP}}/2\pi=267$\,MHz).

The use of a cavity also allows for state detection by differential transmission \cite{boozer:2006, khudaverdyan:2009, reichel:2010} and we now compare the fluorescence and transmission techniques in the same experimental setup. In the regime of small probe laser powers, the cavity transmission is given by $T=1/\left|1-\nu\right|^{2}$. Atomic hyperfine states can be distinguished because $\Delta_a=0$ for an atom in state \mbox{$F=2$} (resonant case, minimum transmission) and $\Delta_a/2\pi\approx6.8$\,GHz for an atom in \mbox{$F=1$} (off-resonant case, high transmission). The experimental sequence for differential transmission is equivalent to the fluorescence measurement, but the probe laser is applied along the cavity axis for 300\,\textmu s and is $\sigma^+$-polarized such that it drives the \mbox{$\left|F=2,m_F=2\right\rangle\leftrightarrow$$\left|F'=3,m_{F'}=3\right\rangle$} cycling transition. A magnetic field of $\approx250$\,mG along the cavity axis provides a quantization axis. We set cavity and probe laser frequencies equal ($\Delta_c=0$) and vary the detuning $\Delta_a$. The measured probe transmission remains at the empty cavity value (100\,\%) with the atom prepared in the \mbox{$F=1$} state. With the atom prepared in the \mbox{$F=2$} state, the transmission reduces to approximately 40\,\% for $\Delta_a=0$\,MHz (Fig.~\ref{fig:detuning}) in agreement with theory. This allows us to discern the hyperfine states with a fidelity of 99.0$\pm$0.5\,\% from a single 300 \textmu s probe interval (quoted error is statistical). While longer probe intervals can theoretically increase the fidelity, this is accompanied by probe light induced heating and atom loss. Moreover, the measured state detection fidelity reduces dramatically when the atomic detuning is increased up to $\Delta_a/2\pi=40$\,MHz (Fig.~\ref{fig:detuning}a).

In comparison, we find that cavity-enhanced fluorescence outperforms differential transmission. It achieves higher fidelities over a large range of detunings while operating at much higher readout speeds. In addition, the required cavity parameters (moderate size, moderate linewidth) are generally easier to implement.

By means of cavity-enhanced fluorescence, we have thus introduced an efficient realization of DiVincenzo's requirement for qubit readout \cite{divincenzo:2000} for neutral atoms. The elimination of atom loss at detection establishes trapped neutral atoms as truly stationary qubits. Our scheme proves robust against atomic detuning allowing operation in deep optical dipole traps. It does not require the strong-coupling regime of cavity quantum electrodynamics which simplifies technical implementation. The combination of experimental robustness and readout speeds which are fast compared to hyperfine qubit decoherence times \cite{weinfurter:2008} makes cavity-enhanced fluorescence state detection a useful tool for quantum protocols based on neutral atoms \cite{kimble:2008, meschede:2009, grangier:2010, saffman:2010}. Using the existing capabilities for deterministic atom transport in optical dipole traps \cite{nussmann:2005, meschede:2006, chapman:2007}, the cavity can serve as a readout head into and out of which neutral atom qubit registers are shifted. Lossless atomic state detection can also improve the performance of atomic clocks \cite{lemonde:2009}. Finally, our scheme is applicable to other physical systems with optically accessible qubits such as trapped ions, quantum dots, diamond NV centers or cold molecules, and can be used to speed up existing protocols as is important for quantum error correction \cite{myerson:2008}.

This work was partially supported by the Deutsche Forschungsgemeinschaft (Research Unit 635) and the European Union (IST projects SCALA and AQUTE). D. L. M. acknowledges support from the Alexander von Humboldt Foundation.


\begin{thebibliography}{10}
\bibitem{kimble:2008}
H. J. Kimble,
\newblock Nature {\bf 453}, 1023 (2008).

\bibitem{kubanek:2009}
A. Kubanek {\em et~al\/}.,
\newblock Nature {\bf 462}, 898 (2009).

\bibitem{meschede:2009}
M. Karski {\em et~al\/}.,
\newblock Science {\bf 325}, 174 (2009).

\bibitem{grangier:2010}
T. Wilk {\em et~al\/}.,
\newblock Phys. Rev. Lett. {\bf 104}, 010502 (2010).

\bibitem{saffman:2010}
L. Isenhower {\em et~al\/}.,
\newblock Phys. Rev. Lett. {\bf 104}, 010503 (2010).

\bibitem{blatt:2004}
C. F. Roos {\em et~al\/}.,
\newblock Phys. Rev. Lett. {\bf 92}, 220402 (2004).

\bibitem{acton:2006}
M. Acton {\em et~al\/}.,
\newblock Quantum Inf. Comput. {\bf 6}, 465 (2006).

\bibitem{hume:2007}
D. B. Hume, T. Rosenband, and D. J. Wineland,
\newblock Phys. Rev. Lett. {\bf 99}, 120502 (2007).

\bibitem{olmschenk:2007}
S. Olmschenk {\em et~al\/}.,
\newblock Phys. Rev. A {\bf 76}, 052314 (2007).

\bibitem{myerson:2008}
A. H. Myerson {\em et~al\/}.,
\newblock Phys. Rev. Lett. {\bf 100}, 200502 (2008).

\bibitem{kuhr:2005}
S. Kuhr {\em et~al\/}.,
\newblock Phys. Rev. A {\bf 72}, 023406 (2005).

\bibitem{volz:2006}
J. Volz {\em et~al\/}.,
\newblock Phys. Rev. Lett. {\bf 96}, 030404 (2006).

\bibitem{yavuz:2006}
D. D. Yavuz {\em et~al\/}.,
\newblock Phys. Rev. Lett. {\bf 96}, 063001 (2006).

\bibitem{jones:2007}
M. P. A. Jones {\em et~al\/}.,
\newblock Phys. Rev. A {\bf 75}, 040301(R) (2007).

\bibitem{lengwenus:2007}
A. Lengwenus {\em et~al\/}.,
\newblock Appl. Phys. B {\bf 86}, 377 (2007).

\bibitem{reichel:2010}
R. Gehr {\em et~al\/}.,
\newblock Phys. Rev. Lett. {\bf 104}, 203602 (2010).

\bibitem{khudaverdyan:2009}
M. Khudaverdyan {\em et~al\/}.,
\newblock Phys. Rev. Lett. {\bf 103}, 123006 (2009).

\bibitem{boozer:2006}
A. D. Boozer {\em et~al\/}.,
\newblock Phys. Rev. Lett. {\bf 97}, 083602 (2006).

\bibitem{nussmann:2005b}
S. Nussmann {\em et~al\/}.,
\newblock Nature Phys. {\bf 1}, 122 (2005).

\bibitem{darquie:2005}
B. Darqui{\'{e}} {\em et~al\/}.,
\newblock Science {\bf 309}, 454 (2005).

\bibitem{orozco:2009}
M. L. Terraciano {\em et~al\/}.,
\newblock Nature Phys. {\bf 5}, 480 (2009).

\bibitem{kozuma:2009}
M. Takeuchi {\em et~al\/}.,
\newblock arXiv:0907.0336v1 [quant-ph] (2009).

\bibitem{murr:2003}
K. Murr,
\newblock J. Phys. B {\bf 36}, 2515 (2003).

\bibitem{bochmann:2008}
J. Bochmann {\em et~al\/}.,
\newblock Phys. Rev. Lett. {\bf 101}, 223601 (2008).

\bibitem{divincenzo:2000}
D. P. DiVincenzo,
\newblock Fortschr. Phys. {\bf 48}, 771 (2000).

\bibitem{weinfurter:2008}
W. Rosenfeld {\em et~al\/}.,
\newblock Phys. Rev. Lett. {\bf 101}, 260403 (2008).

\bibitem{nussmann:2005}
S. Nussmann {\em et~al\/}.,
\newblock Phys. Rev. Lett. {\bf 95}, 173602 (2005).

\bibitem{meschede:2006}
Y. Miroshnychenko {\em et~al\/}.,
\newblock Nature {\bf 442}, 151 (2006).

\bibitem{chapman:2007}
K. M. Fortier {\em et~al\/}.,
\newblock Phys. Rev. Lett. {\bf 98}, 233601 (2007).

\bibitem{lemonde:2009}
J. Lodewyck, P.G. Westergaard, and P. Lemonde,
\newblock Phys. Rev. A {\bf 79}, 061401(R) (2009).

\end{thebibliography}
\end{document}